# A Process Algebra Software Engineering Environment

*Bob Diertens*

Programming Research Group, Faculty of Science, University of Amsterdam

## ABSTRACT

In previous work we described how the process algebra based language PSF can be used in software engineering, using the ToolBus, a coordination architecture also based on process algebra, as implementation model. In this article we summarize that work and describe the software development process more formally by presenting the tools we use in this process in a CASE setting, leading to the PSF-ToolBus software engineering environment. We generalize the refine step in this environment towards a process algebra based software engineering workbench of which several instances can be combined to form an environment.

## 1. Introduction

In [10-12], we investigated the use of process algebra, in particular the process algebra based language PSF (Process Specification Formalism), in the software development process. We used the ToolBus, a process algebra based coordination architecture, as implementation model. We give a description of PSF and the ToolBus in sections 1.2 and 1.3. This work resulted in two libraries for PSF, one for software architecture specification, and one for specifying the software as ToolBus application. We obtain a ToolBus application specification by refining an architecture specification with the use of vertical and horizontal implementation techniques. The components in the ToolBus application specification can be implemented separately and connected to the ToolBus which is controlled by a ToolBus script that can be derived from the ToolBus application specification. This technology has been successfully used in a new implementation of the simulator from the PSF Toolkit and an integrated development environment (IDE) for PSF.

In this paper we summarize the software development process with PSF and give an example. We describe this software development process more formally by presenting the tools we use in the development process in a Computer-Aided Software Engineering (CASE) setting. We then generalize the refine step from architecture specification to ToolBus application specification in this environment towards a process algebra based software engineering workbench.

In the remainder of this section we give a description of CASE technology and the terminology we use, followed by brief descriptions of PSF, and of the ToolBus.

### 1.1 Computer-Aided Software Engineering

Since the early days of developing software, tools are used to assist in the developing process. In the beginning these were the tools provided by the operating system, such as editors, compilers, and debuggers. With the demand for larger software systems, the development process became more complex and expensive, and a need to improve and control the development process arose. One of the technologies to achieve this is computerized applications supporting and (partially) automating software-production activities. This resulted in many tools for all kinds of activities in the software development process, from editing and testing tools to management and documentation tools.

With the growth in development and use of these tools also the terminology to denote the function and activities of these tools increased. This terminology is often confusing or misleading. To reason on CASE technology it is necessary to use a fixed terminology. We use the terminology as proposed by Fuggetta in [18] which has since then been used by many others. The following three categories are used for classifying CASE technology.

Tools
> support individual process tasks.

Workbenches
> support process phases or activities and normally consist of a set of tools with some degree of integration.

Environments
> support at least a substantial part of the software process and normally include several integrated workbenches.

### 1.2 PSF

PSF is based on ACP (Algebra of Communicating Processes) [3] and ASF (Algebraic Specification Formalism) [4]. A description of PSF can be found in [13, 14, 22, 23]. Processes in PSF are built up from the standard process algebraic constructs: atomic actions, alternative composition +, sequential composition ·, and parallel composition ∥. Atomic actions and processes are parameterized with data parameters.

PSF is accompanied by a Toolkit containing among other components a compiler and a simulator that can be coupled to an animation [15]. The tools operate around the tool intermediate language (TIL) [24]. Animations can either be made by hand or be automatically generated from a PSF specification [16]. The animations play an important role in our software development process as they can be used to test the specifications and are very convenient in communication to other stakeholders.

### 1.3 ToolBus

The ToolBus [5] is a coordination architecture for software applications developed at CWI (Amsterdam) and the University of Amsterdam. It utilizes a scripting language based on process algebra to describe the communication between software tools. A ToolBus script describes a number of processes that can communicate with each other and with various tools existing outside the ToolBus. The role of the ToolBus when executing the script is to coordinate the various tools in order to perform some complex task. A language-dependent adapter that translates between the internal ToolBus data format and the data format used by the individual tools makes it possible to write every tool in the language best suited for the task(s) it has to perform.

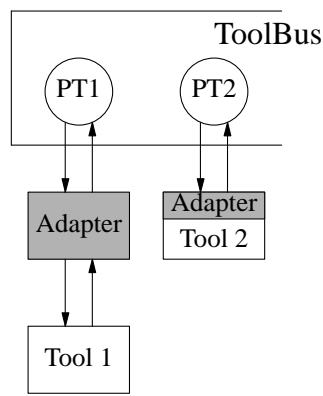

**Figure 1.** Model of tool and ToolBus interconnection

In Figure 1 two possible ways of connecting tools to the ToolBus are displayed. One way is to use a separate adapter and the other to have a builtin adapter. Processes inside the ToolBus can communicate with each other using the actions `snd-msg` and `rec-msg`. ToolBus processes can communicate with the tools using the actions `snd-do` and `snd-eval`. With the latter a tool is expected to send a value back which can be received by the process with the action `rec-value`. A tool can send an event to a ToolBus process which can be received with the action `rec-event`. Such an event has to be acknowledged by the ToolBus process with the action `snd-ack-event`.

## 2. Software Engineering with PSF

A software design consist of several levels, each lower one refining the design on the higher level. The highest level is often referred to as the architecture, the organization of the system as a collection of interacting components. In conventional software engineering processes, the architecture is usually described rather informally by means of a boxes-and-lines diagram. Following a lot of research going on in this area architectural descriptions are becoming more formal, especially due to the introduction of architectural description languages (ADL's). A specification in an ADL can be refined (in several steps) to a design from which an implementation of the system can be built. We have developed a library for PSF to specify software architecture.

In software engineering and reengineering it is common practice to decompose systems into components that communicate with each other. The main advantage of this decomposition is that maintenance can be done on smaller components that are easier to comprehend. To allow a number of components to communicate with each other a so-called coordination architecture will be required. In connection with PSF we will make use of the ToolBus [5] coordination architecture. We have developed a library for PSF to specify ToolBus applications. It does not contain all functionality from the ToolBus but just what is needed for our purposes.

In the following sections we briefly describe the PSF library for specifying software architectures (formalizing the boxes-and-lines diagram), the PSF library for specifying ToolBus applications, and the implementation techniques we use to get from an architecture specification to a ToolBus application specification. We also give an example of how to use the libraries and how to apply the implementation techniques.

### 2.1 Software Architecture Specification

We specify software architecture in PSF with the use of a PSF library providing architecture primitives. The primitives are `snd` and `rec` actions for communication, each taking a connection and a data term as argument. A connection can be built up with a connection function `>>` with two identifiers indicating a component as argument. Processes describing the software architecture with these primitives can be set in an architecture environment, also provided by the PSF library. The architecture environment takes care of encapsulation to enforce the communication between the processes.

### 2.2 ToolBus Application Specification

A specification of a ToolBus application consists of specifications for the processes inside the ToolBus, the ToolBus script, and specifications for the tools outside the ToolBus with which the ToolBus processes communicate. There are two sets of primitives, one set for communications between the processes inside the ToolBus, and one set for communications between the ToolBus processes and the tools. The first set consists of `tb-snd-msg` and `tb-rec-msg` each taking three arguments, the identifier of the sender, the identifier of the receiver, and a term representing a message. The second set consists of the ToolBus process actions `tb-rec-event`, `tb-snd-ack-event`, `tb-snd-do`, `tb-snd-eval`, and `tb-rec-value`, taking a tool identifier and a term as arguments, and the tool actions `tooltb-snd-event`, `tooltb-rec-ack-event`, `tooltb-rec`, `tooltb-snd-value`, taking a single term as argument.

### 2.3 From Architecture to ToolBus Application Design

It is only useful to invest a lot of effort in the architecture if we can relate it to a design on a lower level. In this section we describe the implementation techniques we use to get from an architecture specification to a ToolBus application specification. These techniques are important in our software engineering process for they compress a large number of algebraic law applications into a few steps and so make the process feasible. We demonstrate these techniques with a toy example in section 2.4.

**Horizontal Implementation**

Given two processes $S$ and $I$, $I$ is an implementation of $S$ if $I$ is more deterministic than (or equivalent to) S. As the actions $S$ and $I$ perform belong to the same alphabet, $S$ and $I$ belong to the same abstraction level. Such an implementation relation is called *horizontal*.

To achieve a horizontal implementation we use parallel composition, which can be used to constrain a process. Consider process $P = a \cdot P$, which can do action $a$ at every moment. If we put $P$ in parallel with the process $Q = x \cdot b \cdot Q$ with communication $a \mid b = c$ and enforcing the communication by encapsulation, process $P$ can only do action $a$ whenever process $Q$ has first done action $x$. So process $P$ is constrained by $Q$ and $P \parallel Q$ is an horizontal implementation of $P$, provided $Q$ only interacts with $P$ through $b$. This form of controlling a process is also known as *superimposition* [8] or *superposition* [20].

**Vertical Implementation**

In [32], action refinement is used as a technique for mapping abstract actions onto concrete processes, called *vertical* implementation, which is described more extensively in [33]. With vertical implementation we want to relate processes that belong to different abstraction levels, where the change of level usually comes with a change of alphabet. For such processes we like to develop *vertical* implementation relations that, given an abstract process $S$ and a concrete process $I$, tell us if $I$ is an implementation for the specification $S$. More specifically, we want to develop a mapping of abstract actions to sequences of one or more concrete actions so that $S$ and $I$ are *vertical bisimular*.

We give a rationale of vertical implementation. Consider the processes $P = a \cdot b$ with $a$ an internal action and $Q = c \cdot d \cdot e$ with internal actions $c$ and $d$. If we refine abstract action $a$ from process $P$ to the sequence of concrete actions $c \cdot d$ and rename action $b$ to $e$ we obtain process $Q$. The processes $P$ and $Q$ are vertical bisimular with respect to the mapping consisting of the above refinement and renaming.

We can explain the notion *vertical bisimular* by the following. We hide the internal action $a$ of process $P$ by replacing it with the silent step $\tau$ to obtain $P = \tau \cdot b$. Applying the algebraic law $x \cdot \tau = x$ gives us $P = \tau \cdot \tau \cdot b$. If we now replace the first $\tau$ with $c$ and the second with $d$, and rename $b$ into $e$ we obtain the process $Q$. With $H$ as hide operator and $R$ as renaming operator we can prove that $R_{\{b \to e\}}(H_{\{a\}}(P))$ and $H_{\{c,d\}}(Q)$ are rooted weak bisimular (see Figure 2). So vertical bisimulation is built on rooted weak bisimulation as horizontal implementation relation.

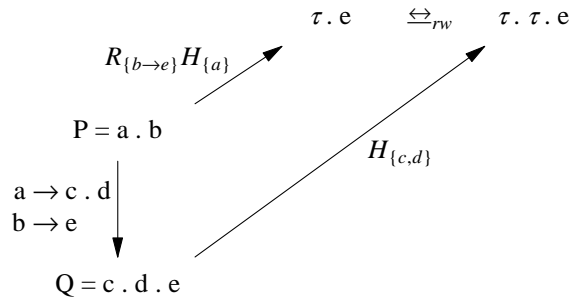

**Figure 2.** Implementation relations

## 2.4 Example

We show our development process for a small application. In this example, Tool1 can either send a `message` to Tool2 and then wait for an acknowledgement from Tool2, or it can send a `quit` after which the application will shutdown.

**Architecture Specification**

We first specify a module for the data and id's we use.

   **data module** `Data`

```
    begin
        exports
        begin
            functions
                message : → DATA
                ack : → DATA
                quit : → DATA
                c1 : → ID
                c2 : → ID
        end
        imports
            ArchitectureTypes
    end Data
```

We then specify the system of our application.

```
    process module ApplicationSystem
    begin
        exports
        begin
            processes
                ApplicationSystem
        end
        imports
            Data,
            ArchitecturePrimitives
        atoms
            send-message
            stop
        processes
            Component1
            Component2
        definitions
            Component1 =
                send-message .
                snd(c1 >> c2, message) .
                rec(c2 >> c1, ack) .
                Component1
            + stop .
                snd-quit
            Component2 =
                rec(c1 >> c2, message) .
                snd(c2 >> c1, ack) .
                Component2
            ApplicationSystem = Component1 ‖ Component2
    end ApplicationSystem
```

The `snd-quit` in the process definition for Component1 communicates with the architecture environment followed by a disrupt to end all processes.

Next, we put the system in the architecture environment by means of binding the main process to the System parameter of the environment.

```
    process module Application
    begin
        imports
            Architecture {
                System bound by [
                    System → ApplicationSystem
                ] to ApplicationSystem
                renamed by [
                    Architecture → Application
                ]
            }
    end Application
```

The generated animation of the architecture is shown in Figure 3. Here, Component1 has just sent a message to Component2, which is ready to send an acknowledgement back.

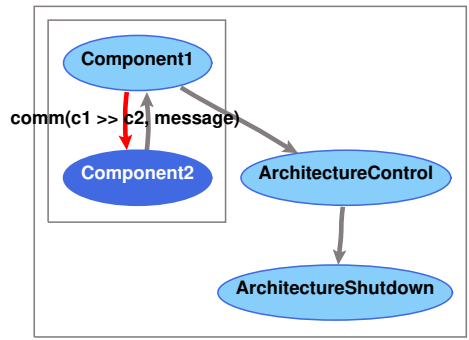

**Figure 3.** Animation of an example architecture

Each box represents an encapsulation of the processes inside the box, and a darker ellipse is a process which is enabled to perform an action in the given state.

The module mechanism of PSF can be used to build more complex components hiding internal actions and sub-processes. With the use of parameterization it is even possible to make several instances of a component.

**ToolBus Application Specification**

We make a ToolBus application specification for our example in the form shown in Figure 1. By refining the specification of the architecture we obtain a ToolBus application specification for our example. Take the process `Component1` from the architecture specification of our toy example.

```
Component1 =
      send-message .
      snd(c1 >> c2, message) .
      rec(c2 >> c1, ack) .
      Component1
   +  stop .
      snd-quit
```

We can make a virtual implementation by applying the mapping consisting of the refinements

```
snd(c1 >> c2, message) → tb-snd-msg(t1, t2, tbterm(message))
rec(c2 >> c1, ack)     → tb-rec-msg(t2, t1, tbterm(ack)) .
                         tb-snd-ack-event(T1, tbterm(message))
snd-quit               → snd-tb-shutdown
```

and the renamings of the local actions

```
send-message           → tb-rec-event(T1, tbterm(message))
stop                   → tb-rec-event(T1, tbterm(quit))
```

Renaming the process `Component1` into `PT1` gives the following result.

```
PT1 =
      tb-rec-event(T1, tbterm(message)) .
      tb-snd-msg(t1, t2, tbterm(message)) .
      tb-rec-msg(t2, t1, tbterm(ack)) .
      tb-snd-ack-event(T1, tbterm(message)) .
      PT1
   +  tb-rec-event(T1, tbterm(quit)) .
      snd-tb-shutdown
```

We can show that `Component1` and `PT1` are vertical bisimular. Applying the renamings on process `Component1` and hiding of the actions to be refined results in

```
Component1' =
      tb-rec-event(T1, tbterm(message)) . τ . τ . Component1'
   +  tb-rec-event(T1, tbterm(quit)) . τ
```

Hiding of the actions in the refinements in process `PT1` results in

```
PT1' =
```

```
        tb-rec-event(T1, tbterm(message)) . τ . τ . τ . PT1'
    +   tb-rec-event(T1, tbterm(quit)) . τ
```

It follows that `Component1'` $\underline{\leftrightarrow}_{rw}$ `PT1'`.

We now make a horizontal implementation by constraining `PT1` with `Tool1Adapter`.

```
    PTool1 = Tool1Adapter ‖ PT1
```

`Tool1Adapter` is itself an constraining of `AdapterTool1` with `Tool1` for which we give the definitions below.

```
AdapterTool1 =
        tooladapter-rec(message) .
        tooltb-snd-event(tbterm(message)) .
        tooltb-rec-ack-event(tbterm(message)) .
        tooladapter-snd(ack) .
        AdapterTool1
    +   tooladapter-rec(quit) .
        tooltb-snd-event(tbterm(quit))
Tool1 =
        snd(message) .
        rec(ack) .
        Tool1
    +   snd(quit)
```

In this constraint, the communication between the actions `tooladapter-rec` and `tooladapter-snd` of `AdapterTool1` and the actions `snd` and `rec` of `Tool1` are enforced.

An implementation for `Component2` can be obtained in a similar way. A generated animation is shown in Figure 4, in which AdapterTool1 just sent a message it had received from Tool1, to ToolBus process PT1.

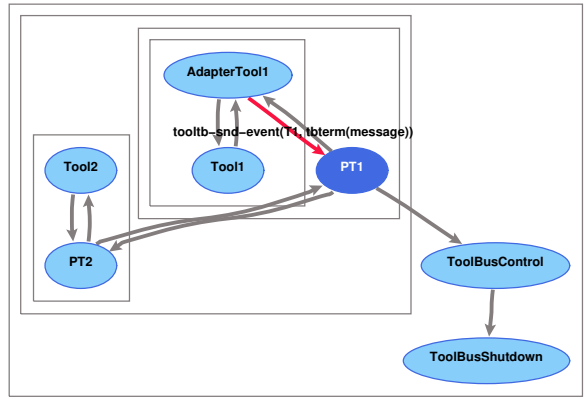

**Figure 4.** Animation of the ToolBus specification example

### Implementation

The implementation consists of three Tcl/Tk [31] programs (Tool1, its adapter, and Tool2), and a ToolBus script. A screendump of this application at work together with the viewer of the ToolBus is shown in Figure 5. With the viewer it is possible to step through the execution of the ToolBus script and view the variables of the individual processes inside the ToolBus. The ToolBus script is shown below. The `execute` actions in the ToolBus script correspond to starting the adapter for Tool1 and starting Tool2 in parallel with the processes `PT1` and `PT2` respectively.

```
process PT1 is
let
    T1: tool1adapter
in
    execute(tool1adapter, T1?) .
    (
        rec-event(T1, message) .
        snd-msg(t1, t2, message) .
        rec-msg(t2, t1, ack) .
```

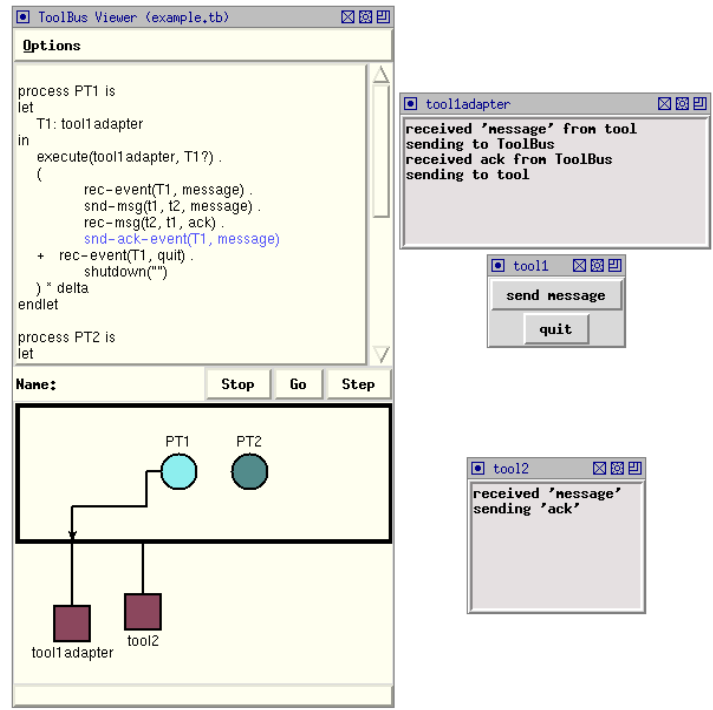

**Figure 5.** Screendump of the example as ToolBus application with viewer

```
        snd-ack-event(T1, message)
   +  rec-event(T1, quit) .
      shutdown("")
   ) * delta
endlet
process PT2 is
let
   T2: tool2
in
   execute(tool2, T2?) .
   (
      rec-msg(t1, t2, message) .
      snd-eval(T2, eval(message)) .
      rec-value(T2, value(ack)) .
      snd-msg(t2, t1, ack)
   ) * delta
endlet
tool tool1adapter is { command = "wish-adapter -script tool1adapter.tcl" }
tool tool2 is { command = "wish-adapter -script tool2.tcl" }
toolbus(PT1, PT2)
```

The processes in the ToolBus script use iteration (*, where P * delta repeats P infinitely) and the processes in the PSF specification use recursion. In PSF it is also possible to use iteration in this case, since the processes have no arguments to hold the current state. On the other hand, in PSF it is not possible to define variables for storing a global state, so when it is necessary to hold the current state, this must be done through the arguments of a process and be formalized via recursion.

Following the description of the ToolBus processes is the description of how to execute the tools by the execute actions. The last line of the ToolBus script starts the processes PT1 and PT2 in parallel.

### 3. The PSF-ToolBus Software Engineering Environment

Development of a software system starts with the specification of its architecture. This architecture specification consists of components that communicate with each other. We make use of the PSF Architecture library that provides the primitives we use in the architecture specification. The system consists of the components put together in parallel. The system is then put in an architecture environment that enforces the communications embedded in the specifications of the components Since the system and the environment are always built in the same way, we can easily generate them. So specification of an architecture is limited to specification of the components. This gives an architecture workbench as shown in Figure 6. Objects to be specified are presented as **bold boxes**, workbench tools as <span style="color:red">ellipses</span>, and generated objects as <span style="color:green">*slanted boxes*</span>.

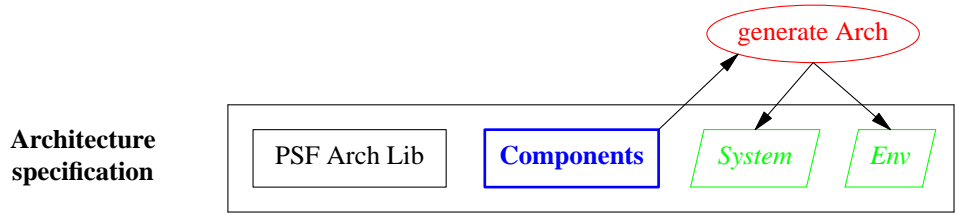

**Figure 6.** The Architecture Workbench

For specification at the ToolBus application level we can make a similar workbench, shown in Figure 7. Again we only need to specify the components, now with the use of the PSF ToolBus library, and the system and environment are again generated.

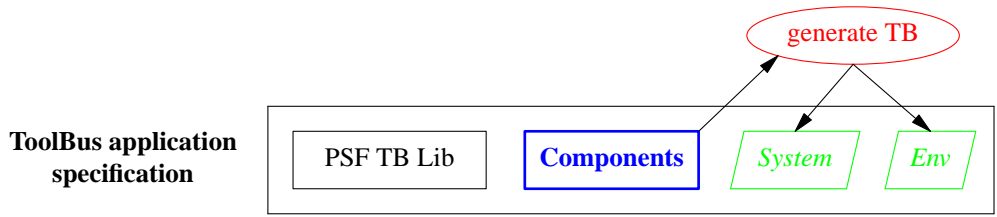

**Figure 7.** The ToolBus Workbench

To derive a ToolBus application specification from an architecture specification we refine the abstract actions in the architecture specification to sequences of actions on the ToolBus application level. The refinement can be done automatically by applying a set of mappings on the specification of the components, resulting in the specification of a set of ToolBus processes. We constrain these ToolBus processes with (abstract) specifications of the tools. The constraining can also be done automatically and results in the specification of the components for the ToolBus application specification.

Combining the workbenches for architecture level and ToolBus level design and integrating the refine and constrain steps, we get the PSF-ToolBus software engineering environment shown in Figure 8. We see that the components for the ToolBus application are generated, so we only have to specify the components on the architecture level and give proper mappings and constraints (the tools) to obtain a ToolBus application specification.

Also shown in Figure 8 is a generation step of a ToolBus-script from the components of the ToolBus application specification. This step is still to be made by hand since PSF specifications use recursion for setting the state of a process, and the ToolBus cannot handle recursive processes. ToolBus scripts use iteration with variable assignment for keeping track of the state of a process.

## 4. A Generalized Process Algebra Software Engineering Environment

The Tools in Figure 8 can also be ToolBus application specifications developed in a similar way. However, the refining of architecture specifications is not limited to the level of ToolBus application specifications. Other levels of design, and even several levels of refinement connected in series are possible.

In the following sections we generalize the concept from the PSF-ToolBus software engineering environment.

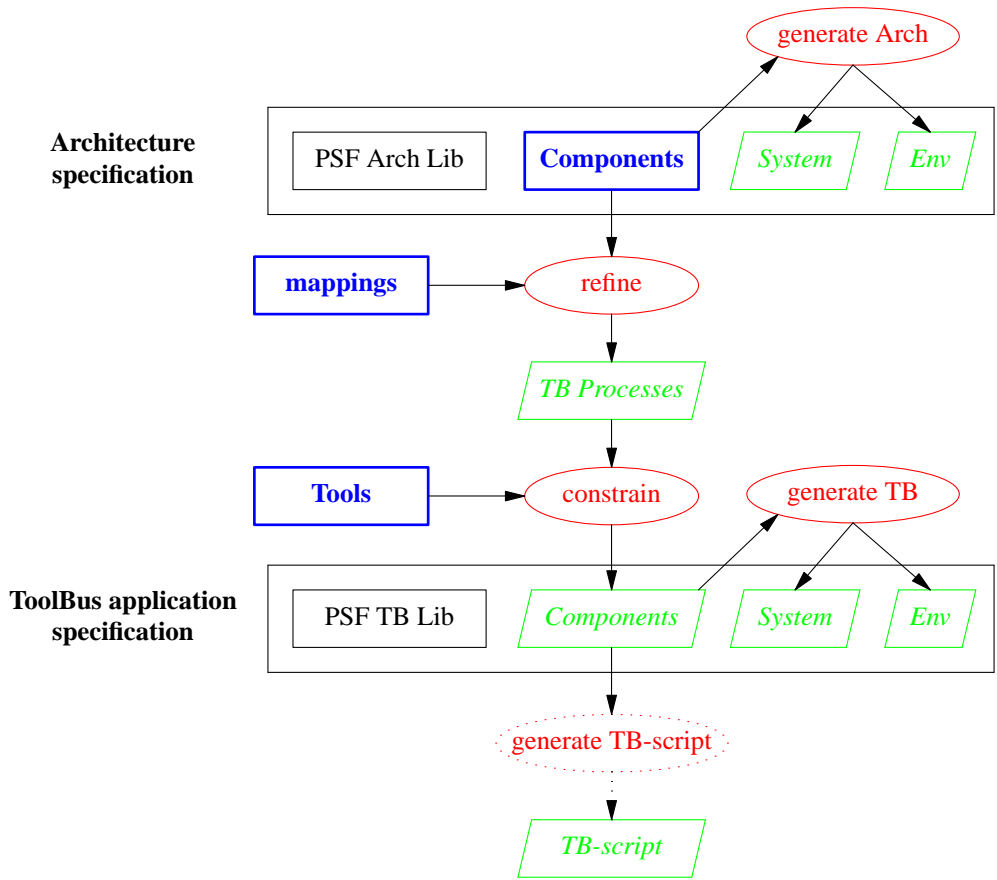

**Figure 8.** The PSF-ToolBus SE Environment

### 4.1 A Generalized PSF Software Engineering Workbench

We can generalize the refine step in the PSF-ToolBus software engineering environment resulting in the workbench shown in Figure 9. The refine and constrain tools are general enough to work on the different level of the design. The generate Level$_X$ tool and PSF Level$_X$ Library can only be applied at level $X$.

### 4.2 A Process Algebra Software Engineering Workbench

So far, we used PSF as process algebra language with the workbenches. However, similar workbenches can be set up for variants of PSF or process algebra based languages similar to PSF. By generalizing from PSF we obtain a Process Algebra Software Engineering Workbench.

If this process algebra language can be translated to TIL, the intermediate language of the PSF Toolkit, the simulator and animation tools from the PSF Toolkit can be used. If another intermediate language (or the process algebra language itself) is used, then a simulator and animation tool for this intermediate language have to be developed.

When using a different intermediate language, reuse of the simulator from the PSF Toolkit is possible. In the implementation of the simulator we only need to replace the kernel with a kernel for the intermediate language, thus reusing the design of the PSF simulator as presented in [11].

### 4.3 Forming an Environment

Several instances of the generalized workbench can be combined to form a software engineering environment. The instances can be connected in series. The specifications of the constraining processes can also be developed using instances of the generalized workbench, leading to an environment in which the workbenches are connected in parallel as well as in series.

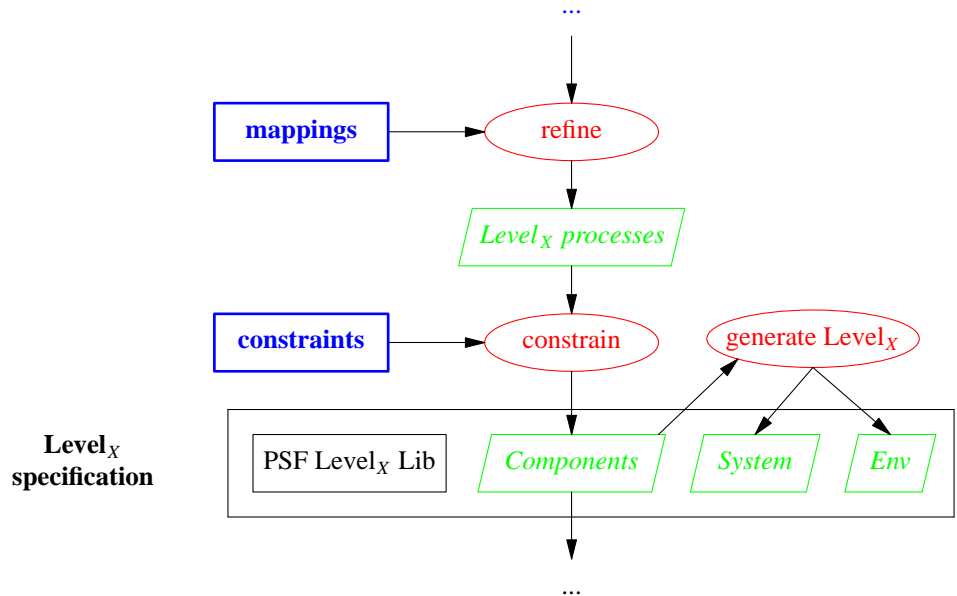

**Figure 9.** The PSF SE Workbench

## 5. Related Work

In the literature several architecture description languages have been proposed and some are based on a process algebra, such as *Wright* [2], *Darwin* [21], and *PADL* [6]. A comparison of several ADL's can be found in [25]. Most of the ADL's do not have any or very little support for refinement. SADL [26][27] however, has been specially designed for supporting architecture refinement. In SADL, different levels of specifications are related by refinement mappings, but the only available tool is a checker. LOTOS [7], a specification language similar to PSF, is used in [19] for the formal description of architectural styles as LOTOS patterns, and in [34] it is used as an ADL for the specification of middleware behavior.

Formal development techniques such as B [1], VDM [17], and Z [9] provide refinement mechanisms, but they do not have support for architecture descriptions. The $\pi$-Method [28] has been built from scratch to support architecture-centric formal software engineering. It is based on the higher-order typed $\pi$-calculus and mainly built around the architecture description language $\pi$-ADL [29] and the architecture refinement language $\pi$-ARL [30]. Tool support comes in the form of a visual modeler, animator, refiner, and code synthesiser.

To our knowledge there is no work done on generalizing software engineering workbenches and creating software engineering environment from instances of the generalized workbenches. There are many meta software development environments with which an environment can be created by integrating a set of existing tools. Such integration can easily be developed with the PSF-ToolBus software engineering environment as is shown in [12]. Here, an integrated development environment for PSF is created from the tools of the PSF Toolkit using the ToolBus to control the communication between the tools.

## 6. Conclusions

We have described the software engineering development process with PSF more formally by presenting the tools we use in the development process in a CASE setting. This resulted in an PSF-ToolBus software engineering environment consisting of two workbenches and a refine step. We generalized the refine step in the environment to an PSF software engineering workbench. Instances of this generalized workbench can be combined to form a software engineering environment. We also generalized from the process algebra based language PSF to obtain a process algebra software engineering environment.

Using several levels for design of software systems has some advantages, of which the most important is that maintenance becomes easier. Adjustments and changes can be made at an appropriate abstract level of design and be worked down the lower levels. The influence of an adjustment or change on design decisions at the lower levels becomes clear in this process, and can be dealt with at the right abstract level. Also, the specifications are part of the documentation of the software system. Simulation of the specifications gives a good understanding of the design of the system, certainly in combination with animations generated from the specifications.

Another advantage is that not only parts of the implementation can be reused for other systems, but also the design can be reused. This is especially useful in an environment where similar products are being made or incorporated in other products.

Using workbenches in the engineering of software systems increases the advantages mentioned above, since it improves the understanding of the design process. It also gives the opportunity to reason about the design process in a more abstract setting.

A disadvantage is that the design process can be more work than strictly necessary for smaller software systems and for systems that are relatively easy to implement. However, software systems that are used over a long period of time need maintenance and evolve into larger and more complex systems. Reengineering the design of such systems is far more work than a durable design right from the start.